\documentclass{article}
\usepackage{graphicx}
\usepackage{url}
\usepackage{hyperref}
\usepackage{amsmath}
\usepackage{stfloats}
\usepackage{xcolor, soul}
\sethlcolor{green}
\begin{document}
\title{\textbf{Adaptive Warden Strategy for Countering Network Covert Storage Channels}}
\author{Mehdi Chourib$^1$, Steffen Wendzel$^{1,2}$, Wojciech Mazurczyk$^{1,3}$}

\date{$^1$ FernUniversit{\"a}t in Hagen, Germany\\
$^2$ Hochschule Worms, Germany\\
$^3$ Warsaw University of Technology, Poland\\
~\\
{\small \textbf{Contact:} \texttt{mehdi.chourib@studium.fernuni-hagen.de,\\ wendzel@hs-worms.de, wojciech.mazurczyk@pw.edu.pl}}}


\maketitle              
\begin{abstract}
The detection and elimination of covert channels are performed by a network node, known as a warden. Especially if faced with adaptive covert communication parties, a regular warden equipped with a static set of normalization rules is ineffective compared to a dynamic warden. However, dynamic wardens rely on periodically changing rule sets and have their own limitations, since they do not consider traffic specifics. We propose a novel adaptive warden strategy, capable of selecting active normalization rules by taking into account the characteristics of the observed network traffic. Our goal is to disturb the covert channel and provoke the covert peers to expose themselves more by increasing the number of packets required to perform a successful covert data transfer. 
Our evaluation revealed that the adaptive warden has better efficiency and effectiveness when compared to the dynamic warden because of its adaptive selection of normalization rules.
\end{abstract}

\section{Introduction}
A confidential data breach may cause significant financial and reputational losses. The Privacy Rights Clearinghouse (PRC) archived more than 9,000 breach events with around 12 billion records between January 2005 and December 2018 \cite{ref_1}. For example, attackers penetrated 77 million PlayStation user accounts, forcing Sony to disconnect their online platform for several weeks in 2011. Breached data included customers' passwords, home addresses, dates of birth, etc. \cite{ref_2}. Three billion user data of Yahoo was breached in 2013. FriendFinder networks were compromised with 412 million accounts in 2016. Facebook 540 million users were breached in 2019. According to the Juniper research, the cost of data breaches exceeded \$150 million in 2020 \cite{ref_3}.

To protect the network perimeter from cyber attacks, most organizations use common security countermeasures such as firewalls, intrusion detection systems, or anti-malware tools \cite{ref_4}. However, current security defenses show their limitations because malware developers are creating increasingly robust malicious code, rendering security controls ineffective. Particularly in situations where information is either infiltrated or exfiltrated using techniques that are not strictly prohibited by existing security policies.
Currently, network traffic is progressively used to hide information (see examples of real-life malware of this kind, e.g., in \cite{ref_5}). In such a case, the \textit{covert channel} becomes an exploit which is used to transfer information violating security policy through the regular communication channels \cite{ref_6} and its sub-type is known as \textit{network covert channel} (NCC). Simmons \cite{ref_7} described the well-known covert channel scenario, i.e., ``prisoners' problem'', and suggested that covert communication techniques can trick the warden to avoid the leakage of information monitored by a third party. Exploiting the network vulnerability became prominent and gained growing attention from both academia and industry \cite{ref_8}.

Data hiding techniques are intended to slightly modify network traffic, permitting secret communications to remain undetected. The alteration of the data transfer can be achieved via \textit{storage} or \textit{timing} covert channels. The former insert secret data into a hidden data carrier, e.g., a field in a header of some network protocol and extract it by reading this carrier by different processes \cite{ref_9,ref_10}. The latter modify time relationships within the communication flow by changing the inter-packet delays or introducing intentional losses \cite{ref_11}. Hence, minor modifications are difficult to detect using both statistical and heuristic approaches. In the literature, there has been some work performed on countering NCC using both regular (i.e., traditional) and dynamic wardens \cite{ref_12}. However, little or no work has investigated the elimination of covert communication based on the observed network traffic's covert channel characteristics.

That is why in this paper we propose a novel concept of an \textit{adaptive} warden that can automatically enable (initially) inactivated rules based on the observation of passing traffic. This approach relies on a dynamic normalization, which evolves over time, considering the necessity to adapt the normalization behavior based on the observed network behavior. 
Considering the above, the main contributions of this work are:

\begin{enumerate}
    \item proposal of a novel adaptive warden to effectively counter storage NCC;
    \item development of an open source proof-of-concept implementation of the adaptive warden\footnote{The code is available at: \url{https://github.com/zcerx/AdaptiveWarden/}};
    \item an in-depth experimental evaluation of the proposed concept to determine its efficacy, effectiveness, and resource consumption (CPU, RAM).
\end{enumerate}

The remainder of the paper is structured as follows. 
Section 2 introduces the background information and discusses related work.
Section 3 describes the threat model as well as the design and proof-of-concept implementation of the adaptive warden. 
The efficiency and resource consumption of the adaptive warden is evaluated in Section 4. 
Finally, Section 5 concludes our research and highlights potential future directions.

\section{Fundamentals and Related Work} 
From the countermeasure's perspective, defending a network from covert channels is typically achieved by using wardens \cite{ref_13, ref_14}. A warden is a single node in a network that intends to unveil, limit, or eliminate any hidden communications \cite{ref_12, ref_15}. 
Clearly, in an ideal case, the primary objective of the warden is to detect the existence of NCCs. However, in practice this is not always possible as there is currently no one-size-fits-all solution capable of efficiently detecting the presence of all NCCs. Thus, wardens typically perform normalization to eliminate identified ambiguities in the network traffic \cite{ref_16}.

Different fundamental types of wardens must be distinguished.
\textit{Passive} wardens do not influence the monitored traffic to detect the secret communication. In contrast, \textit{active} wardens normalize network traffic to eradicate any ambiguities in the protocol header fields, aiming for the removal of covert channels. When a passive or active warden is \textit{stateless}, it does not take previous network traffic packets into consideration. Yet, a state\textit{ful} warden maintains the information about the previously analyzed packets in order to inspect the upcoming traffic \cite{ref_15, ref_17}. Active wardens can be further divided into two categories: While a \textit{proactive} warden can \textit{intentionally issue ‘probes’ (e.g. crafted packets) for known (or unknown) covert channels to provoke responses}, a \textit{reactive} warden performs its actions only after observing actual covert communications \cite{ref_12}.
The identification of covert channels implies detecting a shared resource that could be used as a carrier for the hidden data flow. Detection itself only examines the flow of network packets to reveal any NCC in operation.

Recently, in \cite{ref_12}, it has been demonstrated that a \textit{dynamic} warden, which periodically and randomly shuffles active normalization rules can significantly interfere with the work of the sophisticated \emph{adaptive} NCCs. However, the adaptive covert communication parties are able to detect the warden's normalization strategy, since it uses a limited number of normalization rules and cannot shift them permanently \cite{ref_18}. The main difference between the (typical) regular warden and the adaptive warden is illustrated in Fig.~\ref{fig:Fig1 regular_adaptive_warden}. The concept of the dynamic warden is not just limited to an active and stateless warden, yet it also resolves the limitations of the typical active warden. To implement the proactive behavior and enhance the capabilities of the dynamic warden, the characteristics of the passing traffic should be considered, too. For example, in situations where covert communication parties, i.e., Covert Sender (CS) and Covert Receiver (CR) infer the absence of normalization rules for some specific hiding technique, they can use it to perform hidden communication in an uninterrupted manner. Thus, the dynamic warden cannot normalize the covert communication channel because it neither observes and analyses the incoming traffic nor adjusts its normalization behavior to the observed network traffic. 

That is why further work is needed to propose new concepts that would be free of these limitations. That is why, in this paper, we propose a concept of the adaptive warden.

\section{Attack scenario and the adaptive warden concept}
In this section, we describe the threat model to deal with and the adaptive warden concept. Then, we described the test-bed and proof-of-concept implementations used for simulations.

\subsection{Threat model}
We assume the following attack scenario: through a different attack vector (e.g., via the infected website, social engineering, etc.), an attacker (CS) can access the infected machine in the secured network. All valuable information stored on the compromised device can be discovered and exfiltrated by the CS. However, traffic in this network is monitored and analyzed using various security solutions, e.g., Data Leakage Prevention (DLP) or warden systems. The attacker then uses NCC to exfiltrate confidential data to his external server (CR). To achieve this, CS and CR use the concept of the \textit{adaptive covert communication} as defined in \cite{ref_12}. 

\begin{figure}[!t]
    \centering
    \includegraphics[width=0.7\linewidth]{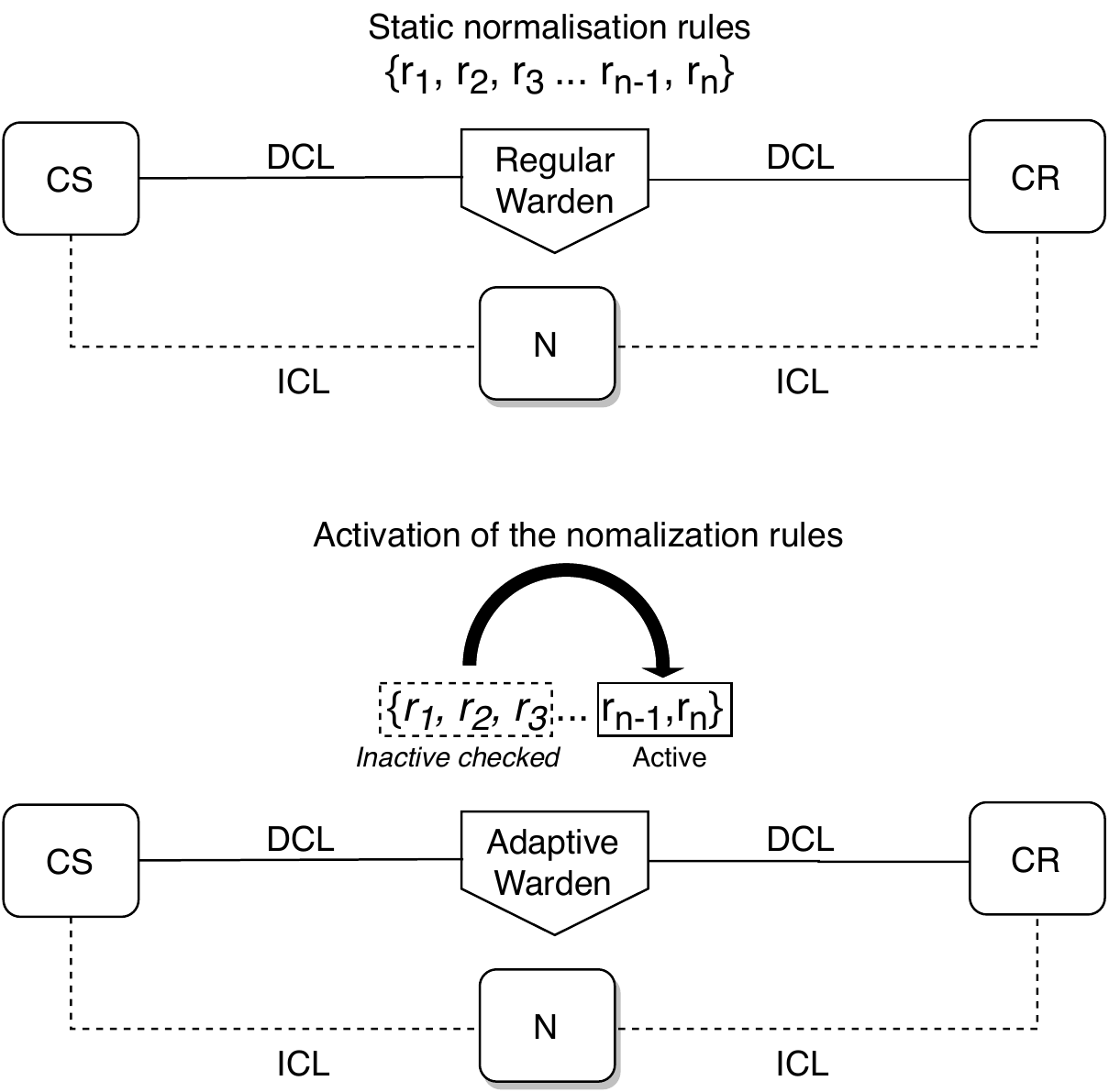}
    \caption{Regular vs.\ adaptive warden.}
    \label{fig:Fig1 regular_adaptive_warden}
\end{figure}

\subsection{Adaptive Covert Communication Functioning}
In the adaptive covert communication, an attacker can infer network protocols, potentially even in the presence of the warden \cite{ref_19}. This scenario assumes that covert communication parties, CS and CR, can either communicate directly (through a direct communication link -- DCL) or indirectly (via an indirect communication link -- ICL). Both links fulfill a specific role in identifying the set of the warden's active normalization rules. On the one hand, DCL is a data link secured by the warden, but only using this communication path it is possible to exfiltrate the desired covert data successfully. Conversely, ICL allows to covertly transfer of very limited information (e.g., metadata) between CS and CR via an intermediate node (N). However, the latter cannot be used for massive transmissions of confidential data. The following section provides more details on how the adaptive covert communication between CS and CR is implemented.

In general, the adaptive covert communication relies on two phases \cite{ref_12}: the network environment learning (NEL) phase and the communication (COM) phase. 
During the NEL phase, the most suitable NCCs are identified by the covert peers to carry out their confidential communication in the presence of an active warden. The CS and CR start by passively monitoring the network traffic. The warden is not performing a NEL phase, only the covert peers do. Once the network protocols are inferred, a handshake process is activated, followed by selecting the suitable network protocols, and packet probes are exchanged to validate whether the warden tolerates the chosen protocol. 

After the NEL phase is initially completed (i.e., when at least one suitable protocol for covert communication is determined), the secret data exchange starts -- the COM phase. Meanwhile, the NEL continues to be active in determining other suitable protocols and detecting any possible modifications introduced by the warden rules that could affect previously established NCCs. When an attacker uses an adaptive covert communication and faces a regular warden, a particular set of static normalization rules can only be applied. Covert parties continuously try different available NCCs, until at least one is found. When this happens, the covert data can be exfiltrated without any further interruptions (due to the static nature of the regular warden). 

In this paper, we assume that if an \textit{adaptive} warden is deployed, it will strive for both covert parties to bypass the warden's active normalization rules by employing the permanent NEL phase and using both DCL and ICL links for efficient communication. It is more challenging since the warden adapts to the NCC flow. 

\subsection{Adaptive Warden}
The concept of an adaptive warden proposed in this paper operates by selecting the rules according to the observed network traffic. During the traffic normalization process, the adaptive warden (which is an active, stateful warden with a proactive behavior) maintains two sets of rules throughout the execution time. The first set is called the \emph{inactive\_checked}, $ic$, and it keeps track of rules that are not applied but evaluated to define whether incoming packets match them. They are a subset of the currently inactive rules, and the $ic$ value ranges between $0$ and $100\%$ of the whole rule set, representing the ratio of the evaluated inactive rules.
The rules from the second set (called \emph{active}) are denoted as $ac$ and defined as a percentage of active rules applied each time. This value is randomly chosen by the warden from the range of $[0\%, 100\%-$ic$]$. The warden using its active rules can either forward, normalize, or drop the inspected network packets.

Based on the observed traffic, the adaptive warden checks and selects its $ac$ rules from the active set (Fig.~\ref{fig:Fig1 regular_adaptive_warden}). It has an initial list of inactive rules and a ratio $(ic)$, which defines how many inactive rules are 'tested' (evaluated); if there is no match, $ic$ rules from the inactive set are evaluated. For each of the matched inactive rules, we register a timestamp (the exact time of a detected activity) called a {\it trigger}. Logged triggers are used later to determine which rules can be moved from the inactive to the active set. We have to track all cases when packets were matching these rules. 

This means that the number of active rules and the time during which they will be applied can be treated as configuration parameters to be tuned to optimize the warden performance. The number of triggers required to relocate an inactive rule is called \emph{threshold\_window\_trigger} ($twt$). The length of the trigger list helps to determine which rule is eligible for the replacement. After a certain period $(twt)$, the initially inactive rules ($ic$ percentage ratio) are moved into the active set. After expiration, all corresponding triggers are discarded. 

\subsection{Experimental setup and proof-of-concept implementations}
The experimental testbed was composed of three hosts: CS, Adaptive Warden, and CR. We assumed that the CS and CR are equipped with adaptive covert communication capabilities. Two network links are available in this experimental setup: DCL and ICL; thus, each host has two network interfaces. As already mentioned, ICL can only be used to share metadata related to exchanged probe packets. On the other hand, the DCL link permits stable covert communications and partially supports the realization of the NEL phase: it keeps sending probe packets through the warden. Moreover, the CS, CR, and the warden run in different virtual machines using Ubuntu 18.04, Intel Dual-Core CPU 2.4 GHz, and 4 GB RAM. 

\subsubsection{Adaptive warden implementation details}
The adaptive warden was implemented in Python, and network traffic normalization was executed using Netfilter firewall (\url{https://mirrors.edge.kernel.org/pub/software/scm/git/git-htmldocs-2.18.1.tar.gz}). The normalization of a specific covert channel has been achieved by the unique corresponding matching rule of the warden,  i.e.,\ each covert packet can be detected by a unique normalization rule of the adaptive warden.
During the initialization, the adaptive warden randomly fills both rule sets and these sets are not disjoint. Moreover, for the rule replacement mechanism, we used two state-of-the-art strategies, i.e., FIFO (First In First Out) and NRU (Not Recently Used). The following rules define their selection:
\textit{(i)} We used FIFO, if none of the rules have been applied before (at the start-up) or if all applied rules were used the same number of times;
\textit{(ii)} We acted in accordance with NRU, if some rules have been already applied.

\subsubsection{Dynamic (and regular) warden implementation details}
To compare the experimental results of the proposed adaptive warden with existing solutions, we used the implementations of the dynamic warden taken from our previous research \cite{ref_12}. In the case of the regular warden, a predefined $R_r$ subset of normalization rules is loaded, and it remains the same during its functioning. On the other hand, the dynamic warden randomly activates a chosen $R_d$ subset of the normalization rules during the length $r_i$. After that, another $\lvert R_d \rvert$ rules were randomly chosen. 

\subsubsection{Adaptive covert communication parties implementation details}
The covert channel traffic was generated with the tool \emph{NELphase} (\url{https://github.com/cdpxe/NELphase}).
During the NEL phase, the CS declares the probe packets sent to the CR through the ICL. The CS randomly selects one network covert channel and transmits to CR a certain number of packets through the DCL. To avoid packet loss during the transmission process, we intentionally send each packet five times. A NCC was considered as not normalized by the warden only if the CR successfully receives the announced test packet during the NEL phase. It means that the communication phase between covert parties could be initiated. The project {\it NEL phase for covert channels} represents a crucial part for correct and robust experiments. 

Suppose that if CR has not received any of the expected packets after a specific time, then in that case, the CS is notified through the ICL that a particular packet did not pass through the warden using this specific covert channel. In other words, announcements and feedback are exchanged between covert peers through the ICL to declare sent probe packets and consequently determine whether a packet reached the destination (feedback channel). CS then pauses for one second before it starts to check the next randomly chosen covert channel. During the NEL phase, the warden continuously updates the list of possible covert channels.

The communication phase starts once both covert parties have detected at least one available covert channel, which currently the warden is not able to normalize. Then, CS sends a sequence of five covert packets via DCL, using the established covert channel. Subsequently, the next nonblocked information hiding technique is selected and used to embed secret data into the packet. This process continues until at least one nonblocked hiding technique remains.

To better analyze the performance of the adaptive warden, we implemented counters in the CR and warden. During the communication phase, CR uses an incremental counter that measures the number of received packets. At the warden level, a counter is used to monitor the number of normalized and forwarded packets along with the measurement of memory and CPU load consumption.

In the proof-of-concept implementation for CS, we used a Scapy tool (\url{https://github.com/secdev/scapy/releases/tag/v2.4.0})
to create and send probe packets that were captured by the CR using Libpcap filters
(\url{https://www.tcpdump.org/release/libpcap-1.9.0.tar.gz}). The adaptive communication parties used a randomly selected technique among the 51 implemented NCC techniques to hide secret data within the probe packets. However, all NCC techniques used for simulation were embedding secret data within various fields of the network packet headers. The communication protocols used in our proof-of-concept were limited to: HTTP, TCP, IPv4, UDP, SCTP, and ICMP.

\begin{figure}[ht]
    \centering
    \includegraphics[width=0.98\textwidth]{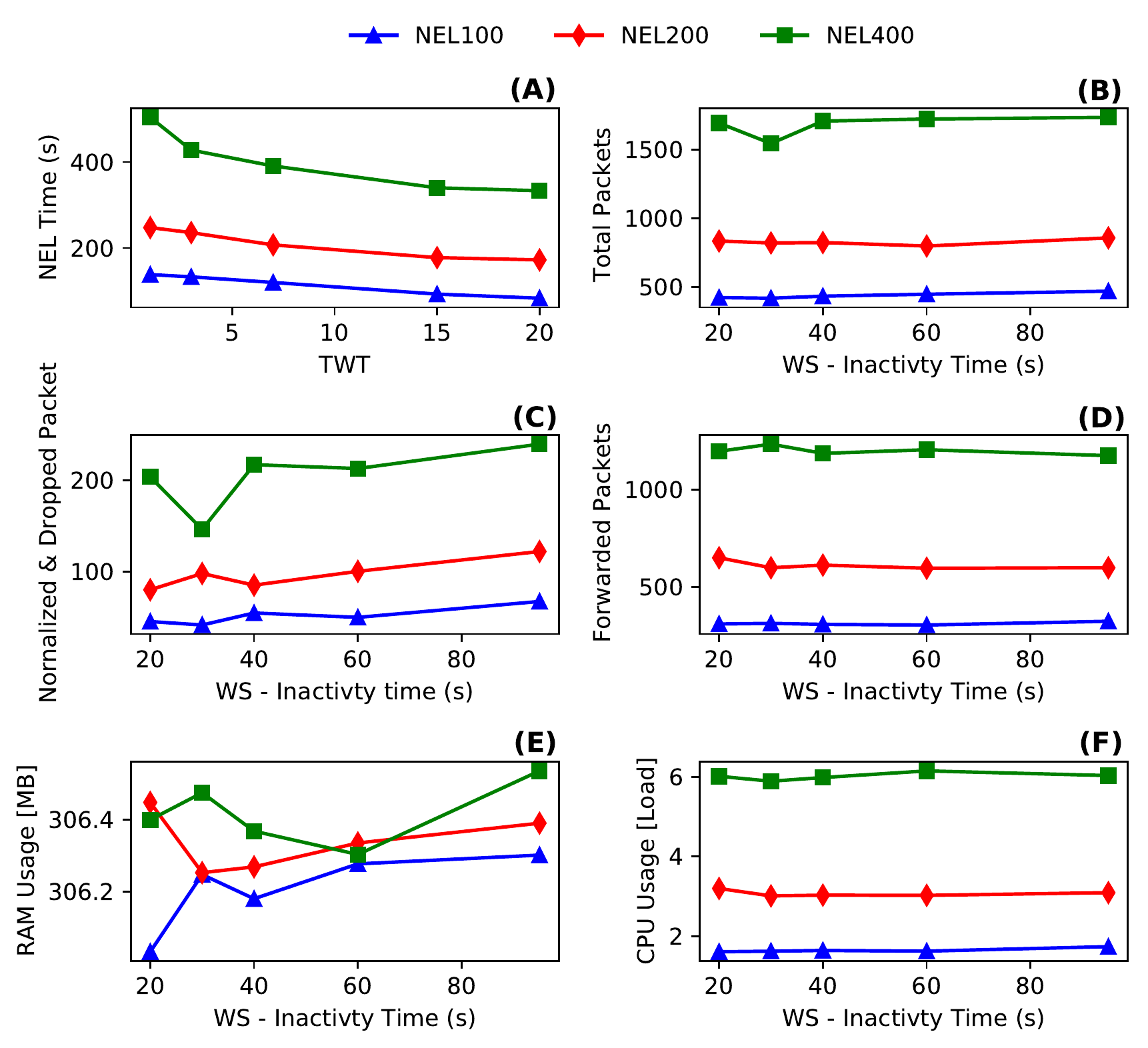}
    \caption{Influence of the: $twt$ on the time needed to complete the transfer of 400 covert packets (A); $ws$ on the total number of packets (B); number of normalized packets (C); forwarded packets (D); RAM usage (E); CPU usage (F)}
    \label{fig:Fig3.pdf}
\end{figure}

\subsubsection{Experimental methodology}
In our evaluation, we assumed that along with the covert data exchange between parties, the NEL phase can be performed simultaneously since the significant goal of the CS and CR is to recognize which active rules are employed by the active warden.

We initiated the measurement immediately after the CR receives the information through the ICL about the first established NCC. During the communication phase, we also recorded the time needed to determine nonblocked covert channel techniques and the time which was needed to transfer a certain number of packets with hidden data (e.g., 400 packets). The measurement ends when the CR receives the declared number of packets. During the experimental evaluation, we assessed the adaptive warden through the variation of the parameters $twt$, $ic$, and $ws$. For each parameter, the experiment was executed three times and the averaged results are presented in the following section.

\section{Results}
In this section, first we conduct an effectiveness analysis of the adaptive warden, then we present the results of the resource consumption. Finally, we compare the performance of the adaptive warden with the previously proposed dynamic warden.

\subsection{Effectiveness analysis}
To evaluate the effectiveness of the adaptive warden, we studied the impact of the percentage (ratio) of the inactive and checked rules ($ic$), values that affect triggers (a delay to enable inactive rules $twt$), and their applicability time $ws$ when all triggers will be discarded.

To present the results, we decided to be coherent and placed a focus on the window size ($ws$) parameter as the most crucial factor between all parameters affecting performance. Moreover, we also inspected how $ws$ influences the time needed for covert communication parties (CS and CR) to transmit 100 (NEL100), 200 (NEL200), and 400 (NEL400) packets with hidden data. 

The transfer of 400 covert packets under the adaptive warden with a single trigger and a high number of evaluated inactive rules takes much longer in comparison to more than one trigger and a few inactive rules (see Fig.~\ref{fig:Fig3.pdf}A). The adaptive attribute of the warden swiftly activated the rules based on the observed network packets and imposed the adaptive covert sender to change covert channel techniques frequently. 
In the result, the hidden transfer required more time and was potentially easier to spot. In other words, it is more difficult to find a nonblocked NCC for an attacker, and thus a defender has more opportunities to ``spot'' the hidden data exchange. 

\begin{figure}[!t]
    \centering
    \includegraphics[width=0.99\textwidth]{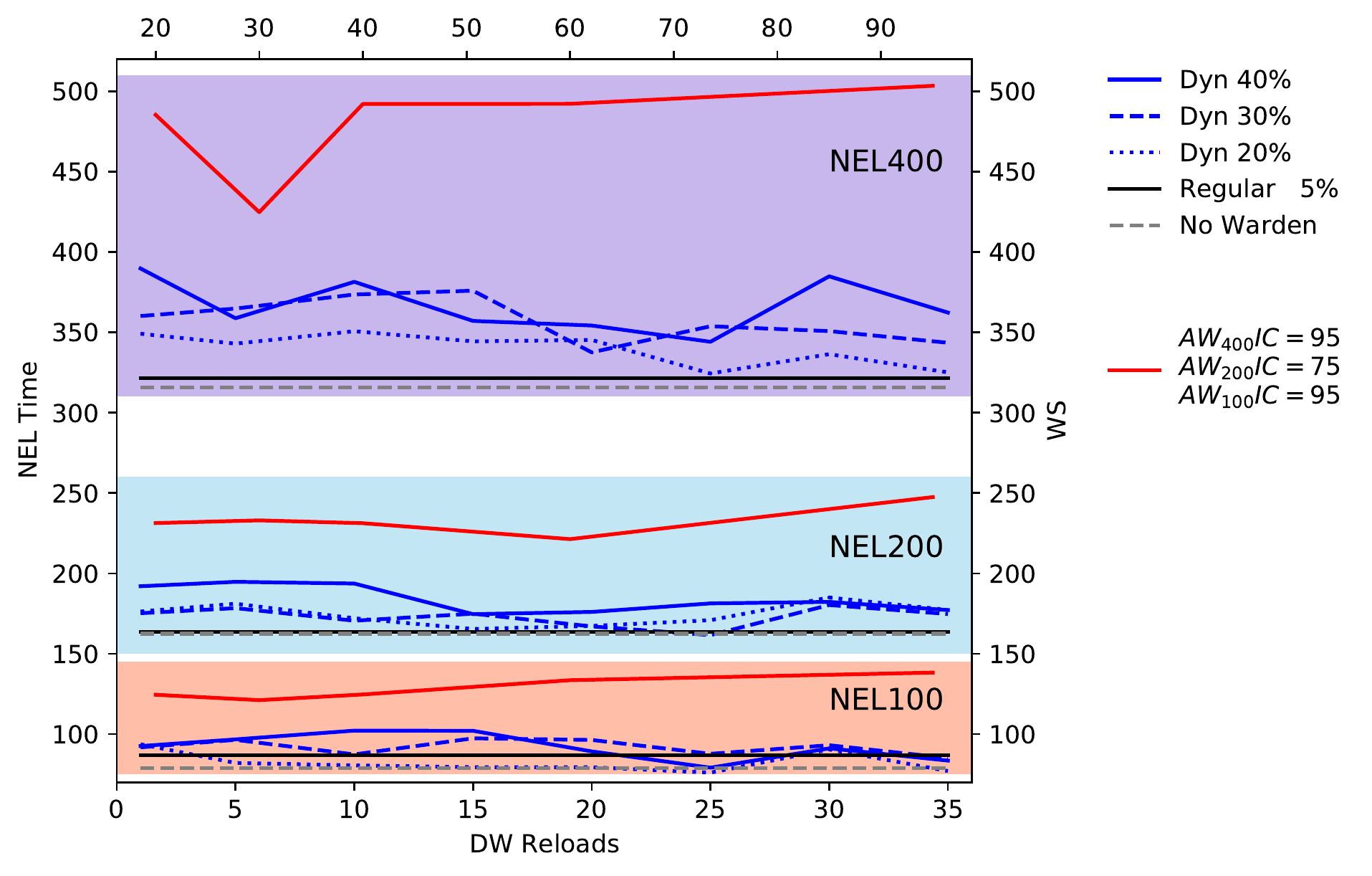}
    \caption{NEL time comparison between scenarios: dynamic vs.\ adaptive warden}
    \label{fig:Fig4 NEL time comparative}
\end{figure}

To continue the evaluation of the warden's effectiveness, the simulation parameters were modified as follows. The number of triggers for moving a rule from the inactive to active set is \textit{twt = \{1, 3, 7, 15, 20\}}, while the fraction of the inactive rules equals \textit{ic = \{1\%, 3\%, 10\%, 25\%, 50\%, 75\%, 95\%\}}. Finally, the time for which inactive rules remain inactive has been set to \textit{ws = \{20s, 30s, 40s, 60s, 95s\}}.

The results presented in Fig.~\ref{fig:Fig3.pdf}B demonstrate that the adaptive warden yields the best results for the highest set of initially inactive rules $(ic = 95\%)$. The lowest trigger $(twt = 1)$ moved the rules from inactive to active for the maximum number of seconds $(ws = 95s)$. It was the maximum period of time when the trigger was applied. Moreover, the results were approximately $50\%$ better when compared to the case when the time needed to transfer the covert data for the scenario of the adaptive warden with $twt = 20$ triggers. The longest period needed to transfer 400 packets with hidden data is when the values of $ic$ and $ws$ were high, while $twt$ was low. In other words, it means a fast reaction, a wide set of possible countermeasures, and long memory.

By blocking some NCC attempts, the warden obliges the covert peers to try another information hiding technique. Since both the detection and normalization are operated immediately, the next inactive rules will be engaged. This complicates the covert communication even more and increases the chance to block NCCs. Rules are active while they are triggered by repetitive attempts of covert communication. 

We also examined the network traffic statistics for the adaptive warden, particularly the rate of network packets: normalized and dropped, forwarded without alteration, and transferred.
Fig.~\ref{fig:Fig3.pdf}B, C, and D showcase that the adaptive warden provides the highest performance against the adaptive covert communication scenario when configured with a high ratio $ic$ for the inactive set of rules and low $twt$ and $ws$ values. 
These results also illustrate that the adaptive warden with a single trigger ($twt=1$) dropped ca.\ 5 times more packets, forwarded 5\% fewer packets, and processed 36\% more packets, compared to the one with 20 triggers ($twt = 20$). The fastest warden reaction is the best strategy to introduce obstacles in the NCC traffic. 

Nearly immediate response to the detected covert packets makes the investigation meaningless for possible or recommended probability distributions. This is applicable only for dynamic wardens because they switch rules randomly. Note that the proposed approach aims at countering the frequent change of the network covert channels used by covert parties.
Moreover, it means that the proposed adaptive warden has a deterministic nature, since it acts when a covert sender's attempt is suspected (depends on the real data flow).

Logically, the larger \emph{inactive} rules set provides the higher the chance that at least one rule will be triggered several times and will be moved to the $as$. Moreover, when $twt$ was low, moving rules from the \emph{inactive} to $as$ was likely to happen more frequently. Finally, the auxiliary warden components, such as the time required for thread synchronization, increase the computational delay due to a smaller $ws$.

\subsection{Resource consumption}
Furthermore, the analyses performed on the resource consumption of the adaptive warden regarding CPU and memory usage are illustrated in Figs.\ \ref{fig:Fig3.pdf}E and F. The variation of the parameters, i.e., $twt$, $ic$, and $ws$, affected the internal RAM distribution, while its overall amount remained practically constant. At the same time, the most effective adaptive warden caused the highest CPU workload of 25\%.

\subsection{Comparison of the adaptive warden with dynamic warden}
The results of the comparative analysis of the adaptive warden with the dynamic one are presented in Fig.~\ref{fig:Fig4 NEL time comparative}.
The results indicate that the time needed to transfer 400 covert packets from the CS to the CR using the adaptive warden is 29\% higher than for the dynamic warden. Therefore, the adaptive warden is the most effective measure to fight against adaptive hidden data transfer.
Consequently, the covert communication parties were obliged to transfer more packets in comparison to the dynamic warden. The adaptive warden with $twt = 1$, $ic = 95\%$ and $ws = 95s$, representing the fastest response strategy (dropped 1,736 packets overall). In the case of the dynamic warden, there were ca.\ 1,450 packets dropped, respectively. Results show that the adaptive warden increased the number of packets needed to perform successful covert communications up to ca.\ 58\% compared to the dynamic warden. The adaptive warden mentioned above slows down the data breach significantly. The RAM usage remains constant (306 MB), while the CPU consumption increased as the number of triggers for a certain inactive rule decreased.

To summarize, in our study, we found that for:
\begin{itemize}
  \item \textbf{NEL100}: The NEL time with the adaptive warden was 43\% longer, needed 38\% more packets to complete the hidden data transfer. The adaptive warden normalized and dropped 82\% more packets compared to the dynamic warden; with 6\% more forwarded packets.
  \item \textbf{NEL200}: It is visible that the NEL time with the adaptive warden was 35\% longer, needed 34\% more packets, normalized and dropped 69\% more packets than the dynamic warden with a similar level of forwarded packets.
  \item \textbf{NEL400}: The NEL time with the adaptive warden was 29\% longer, needed 23\% more packets, normalized and dropped 77\% more packets than the dynamic warden, with 29\% more forwarded packets.
\end{itemize}

Regarding CPU load and RAM consumption, for NEL100/200/400, the adaptive warden experienced 82\%, 69\%, and 77\% more CPU loads, respectively, while its RAM utilization remains on par with the dynamic warden. However, the CPU load of the adaptive warden is nearly doubled when compared to the dynamic warden (91\%), but this is caused by the frequent changes of working normalization rules.

Moreover, to complete the hidden data exchange with the adaptive warden, the adaptive covert parties needed ca.\ 1,736 packets versus ca.\ 1,344 packets with the dynamic warden (29\% increase). The adaptive warden is capable of normalizing many more packets (75\%) than the dynamic warden because it adapts its strategy (set of active rules) to the passing traffic. In contrast, the dynamic warden selects them randomly.
Thus, the adaptive warden can be considered as an effective solution: it takes more time to complete the hidden data transfer when compared with the traditional dynamic warden scenario.

\paragraph*{Warden Simulation}
In addition, we simulated all three warden types using the NELphase tool v0.4.0 via WiFi. The average results for the above-mentioned NEL100 scenario using 50\% blocked rules were as follows (please note that the simulations' performance was lower than the one in the actual testbed since sending a packet took longer, which increased the overall time consumption): No warden: 110 sec, regular warden: 141 sec (+28\%), dynamic warden with 10 sec reload interval: 216 sec (+96\%), adaptive warden with 10s reload interval and 10\% inactive-checked rules: 343 sec (+212\%).

\section{Conclusions and future work}
Information hiding techniques that exfiltrate data are becoming more advanced and sophisticated in terms of features and less visible in a general sense. Thus, it is necessary to develop proper countermeasures.  
In this paper, we present the practical evidence that the adaptive warden approach offers good protection against adaptive storage covert channels. Moreover, we shared a proof-of-concept implementation with the community and evaluated the adaptive warden's performance in different setups. 
Our evaluation has shown that the adaptive warden with a single trigger (nearly immediate response to possible NCC) and a high number of inactive rules (wide scope of possible counter measures) increased the number of packets needed to perform a successful covert communication compared to the state-of-the-art solution (the \textit{dynamic warden} scenario). 
%
%
In future work, we are planing to further optimize the configuration and performance of the adaptive warden. 


\section*{Acknowledgment}
This work has been partially supported by the SIMARGL project, with the support of the EC under the Horizon 2020 Program, under GA No. 833042.

\end{document}